\documentclass[letter,longauth,traditabstract]{aa}  

\usepackage{graphicx}
\usepackage{txfonts}
\usepackage{natbib}
\def\mcfost{{\sf MCFOST}}
\def\ProDiMo{{\sf ProDiMo}}

\begin{document}

\title{The \emph{Herschel} view of Gas in Protoplanetary Systems (GASPS)\thanks{\emph{Herschel} is an ESA space observatory with
science instruments provided by European-led Principal Investigator
consortia and with important participation from NASA.}} 
\subtitle{First comparisons with a large grid of models}

\author{C.~Pinte\inst{1,2}
  \and
  P.~Woitke\inst{3,4,5},
  F.~M\'enard\inst{1},
  G.~Duch\^ene\inst{6,1},
  I.~Kamp\inst{7},
  G.~Meeus\inst{8},
  G.~Mathews\inst{9},  
  C.D.~Howard\inst{10}, 
  C.A.~Grady\inst{11}, 
  W.-F.~Thi\inst{3,1},
  I.~Tilling\inst{3},
  J.-C.~Augereau\inst{1},
  W.R.F.~Dent\inst{12,13} 
  \and
  J.~M.~Alacid\inst{14,15}, 
  S.~Andrews\inst{16}, 
  D.R.~Ardila\inst{17},
  G.~Aresu\inst{7}, 
  D.~Barrado\inst{18,19}, 
  S.~Brittain\inst{20},
  D.R.~Ciardi\inst{21}, 
  W.~Danchi\inst{22}, 
  C.~Eiroa\inst{8}, 
  D.~Fedele\inst{8,23,24}, 
  I.~de~Gregorio-Monsalvo\inst{12,13}, 
  A.~Heras\inst{25},
  N.~Huelamo\inst{19}, 
  A.~Krivov\inst{26}, 
  J.~Lebreton\inst{1},
  R.~Liseau\inst{27}, 
  C.~Martin-Za\"idi\inst{1}, 
  I.~Mendigut\'ia\inst{19}, 
  B.~Montesinos\inst{19}, 
  A.~Mora\inst{28},
  M.~Morales-Calderon\inst{29}, 
  H.~Nomura\inst{30}, 
  E.~Pantin\inst{31},
  I.~Pascucci\inst{32}, 
  N.~Phillips\inst{3}, 
  L.~Podio\inst{7},
  D.R.~Poelman\inst{5}, 
  S.~Ramsay\inst{33}, 
  B.~Riaz\inst{32},
  K.~Rice\inst{3}, 
  P.~Riviere-Marichalar\inst{19},
  A.~Roberge\inst{34}, 
  G.~Sandell\inst{34},
  E.~Solano\inst{12,13}, 
  B.~Vandenbussche\inst{35},
  H.~Walker\inst{36},
  J.P.~Williams\inst{9}, 
  G.J.~White\inst{36,37}
  \and
  G.~Wright\inst{4}
}

\institute{
 Universit\'e Joseph-Fourier - Grenoble\,1/CNRS, Laboratoire
 d'Astrophysique de Grenoble (LAOG) UMR 5571, BP 53, 38041 Grenoble
 Cedex 09, France\\
  \email{christophe.pinte@obs.ujf-grenoble.fr} 
  \and 
  School of Physics, University of Exeter, UK
  \and 
  SUPA, Institute for Astronomy, University of Edinburgh, Royal Observatory, Blackford Hill, Edinburgh, EH9 3HJ, UK;
  \and 
  UK Astronomy Technology Centre, Royal Observatory, Edinburgh, Blackford Hill, Edinburgh EH9 3HJ, UK
  \and 
  School of Physics \& Astronomy, University of St.~Andrews, North Haugh, St.~Andrews KY16 9SS, UK
  \and 
  Astronomy Department, University of California, Berkeley, CA 94720-3411, USA
  \and 
  Kapteyn Astronomical Institute, Postbus 800, 9700 AV Groningen, The Netherlands
  \and 
  Dep. de F\'isica Te\'orica, Fac. de Ciencias, UAM Campus Cantoblanco, 28049 Madrid, Spain
  \and 
  Institute for Astronomy, University of Hawaii at Manoa, Honolulu, HI 96822, USA
  \and 
  SOFIA-USRA, NASA Ames Research Center, Mailstop 211-3 Moffett Field CA 94035 USA
  \and 
  Eureka Scientific and Exoplanets and Stellar Astrophysics Lab, NASA Goddard Space Flight Center, Code 667, Greenbelt, MD, 20771, USA
  \and 
  ALMA, Joint ALMA Office, Avda Apoquindo 3846, Piso 19, Edificio
  Alsacia, Las Condes, Santiago, Chile
  \and 
  European Southern Observatory, Alonso de Córdova 3107, Vitacura, Casilla 19001, Santiago 19, Chile
  \and 
  Unidad de Archivo de Datos, Depto. Astrof{\'i}sica, Centro de Astrobiolog{\'i}a (INTA-CSIC), P.O. Box 78, E-28691 Villanueva de la Ca\~nada, Spain
  \and 
  Spanish Virtual Observatory
  \and 
  Harvard-Smithsonian Center for Astrophysics, 60 Garden St., Cambridge, MA, USA
  \and 
  NASA Herschel Science Center, California Institute of Technology, Pasadena, USA. 
  \and 
  Calar Alto Observatory, Centro Astron\'omico Hispano-Alem\'an
  C/Jes\'us Durb\'an Rem\'on, 2-2, 04004 Almer\'{\i}a, Spain
  \and 
  LAEX, Depto. Astrof{\'i}sica, Centro de Astrobiolog{\'i}a (INTA-CSIC), P.O. Box 78, E-28691 Villanueva de la Ca\~nada, Spain
  \and 
  Clemson University
  \and 
  NASA Exoplanet Science Institute/Caltech 770 South Wilson Avenue, Mail Code: 100-22, Pasadena, CA USA 91125
  \and 
  NASA Goddard Space Flight Center, Exoplanets \& Stellar Astrophysics, Code 667, Greenbelt, MD 20771, USA
  \and 
  Max Planck Institut f{\"u}r Astronomie, K{\"o}nigstuhl 17, 69117 Heidelberg, Germany
  \and 
  ohns Hopkins University Dept. of Physics and Astronomy, 3701 San Martin drive Baltimore, MD 21210 USA
  \and 
  Research and Scientific Support Department-ESA/ESTEC, PO Box 299, 2200 AG Noordwijk, The Netherlands
  \and 
  Astrophysikalisches Institut und Universit{\"a}tssternwarte, Friedrich-Schiller-Universit{\"a}t, Schillerg{\"a}{\ss}chen 2-3, 07745 Jena, Germany
  \and 
  Department of Radio and Space Science, Chalmers University of Technology, Onsala Space Observatory, 439 92 Onsala, Sweden
  \and 
  ESA-ESAC Gaia SOC, P.O. Box 78. E-28691 Villanueva de la Ca\~nada, Madrid, Spain
  \and 
  Spitzer Science Center, California Institute of Technology, 1200 E California Blvd, 91125 Pasadena, USA.
  \and 
  Department of Astronomy, Graduate School of Science, Kyoto University, Kyoto 606-8502,Japan
  \and 
  CEA/IRFU/SAp, AIM UMR 7158, 91191 Gif-sur-Yvette, France
  \and 
  Space Telescope Science Institute, 3700 San Martin Drive, Baltimore, MD 21218, USA
  \and 
  European Southern Observatory, Karl-Schwarzschild-Strasse, 2, 85748 Garching bei M\"unchen, Germany.
  \and 
  Exoplanets and Stellar Astrophysics Lab, NASA Goddard Space Flight
  Center, Code 667, Greenbelt, MD, 20771, USA 
  \and 
  Instituut voor Sterrenkunde, KU Leuven, Celestijnenlaan 200D, 3001 Leuven, Belgium
  \and 
  The Rutherford Appleton Laboratory, Chilton, Didcot, OX11 OQL, UK
  \and 
  Department of Physics \& Astronomy, The Open University, Milton Keynes MK7 6AA, UK
}

   \date{Received ; accepted }

\abstract{The \emph{Herschel} GASPS Key Program is a survey of the gas
phase of protoplanetary discs, targeting 240 objects which cover a large
range of ages, spectral types, and disc properties. 
To interpret this
large quantity of data 
and initiate self-consistent analyses of the
gas and dust properties of protoplanetary discs, 
we have combined the
capabilities of the radiative transfer code \mcfost\ with the gas
thermal balance and chemistry code \ProDiMo\ to compute a grid of
$\approx$ 300\,000 disc models (DENT).
We present a comparison of the first \emph{Herschel}/GASPS line and continuum
data with the predictions from the DENT grid of models. 
Our objective
is to test some of 
the main trends already identified in the DENT grid, as well as to define
better empirical diagnostics to estimate the total gas mass of
protoplanetary discs. 
Photospheric UV radiation appears to be the dominant gas-heating
mechanism for Herbig stars, whereas UV excess and/or X-rays emission
dominates for T~Tauri stars.
The DENT grid reveals 
the complexity in the analysis of far-IR lines and the difficulty to
invert these observations into physical quantities. 
The combination of
\emph{Herschel} line observations with continuum data and/or with
rotational lines in the (sub-)millimetre regime, in particular CO
lines, is required for a detailed characterisation of the physical and
chemical properties of circumstellar discs.}

\maketitle


\section{Introduction}

The dust phase of circumstellar discs has received a lot of attention
in the last few decades, giving us a clearer picture of their
structure and dust content through many 
continuum surveys in various wavelengths regimes
\citep[e.g.][]{Beckwith90,Andrews07,Evans07bis}, complemented by detailed studies of
individual objects, 
combining spectral energy distributions (SEDs) and resolved maps in scattered light and thermal
emission \citep[e.g.][]{Pinte08b,Duchene10}.

Although  gas represents 99\,\% of the initial mass of discs, 
it has been more difficult to observe and is mostly restricted to millimetre lines
probing the cold outer disc, where the freeze-out of molecules is
important \citep[e.g.][]{Dent05,Schaefer09}, and near-IR lines 
which are only emitted from the hot inner parts of discs \citep[e.g.][]{Najita03,Brittain07}. 
The high sensitivity of \emph{Herschel}
\citep{Pilbratt10} opens an opportunity to
systematically probe
the gas phase of discs, in particular the warm atomic and molecular layer
responsible for the bright gas emission lines in the far-IR.
The GASPS open time key program (see Dent et al., in prep. and \citealp{Mathews10})
is a large survey of gas in discs with a gas mass
sensitivity comparable to the dust surveys. 
GASPS will observe several atomic and
molecular lines in about 240 protoplanetary  disc systems with
ages in the critical 1 to 30 million year age range during which planets
form and the gas seems to dissipate. 

 The interpretation of gas observations is complicated by the large
number of  processes at play: processing of radiation by dust
grains, disc thermal structure, 
chemistry, excitation and destruction of molecules, freeze-out and
desorption on the dust grains, etc. To estimate the relative
importance of these mechanisms as a function of age, stellar
properties, disc structure, and dust content, we have computed a large
grid of synthetic SEDs and gas emission lines, named Disc Evolution with Neat Theory (DENT, \citealp{Woitke10}).
Here, we confront the trends identified in the DENT grid
with the first GASPS observations.

\begin{figure*}
  \includegraphics[width=0.46\hsize]{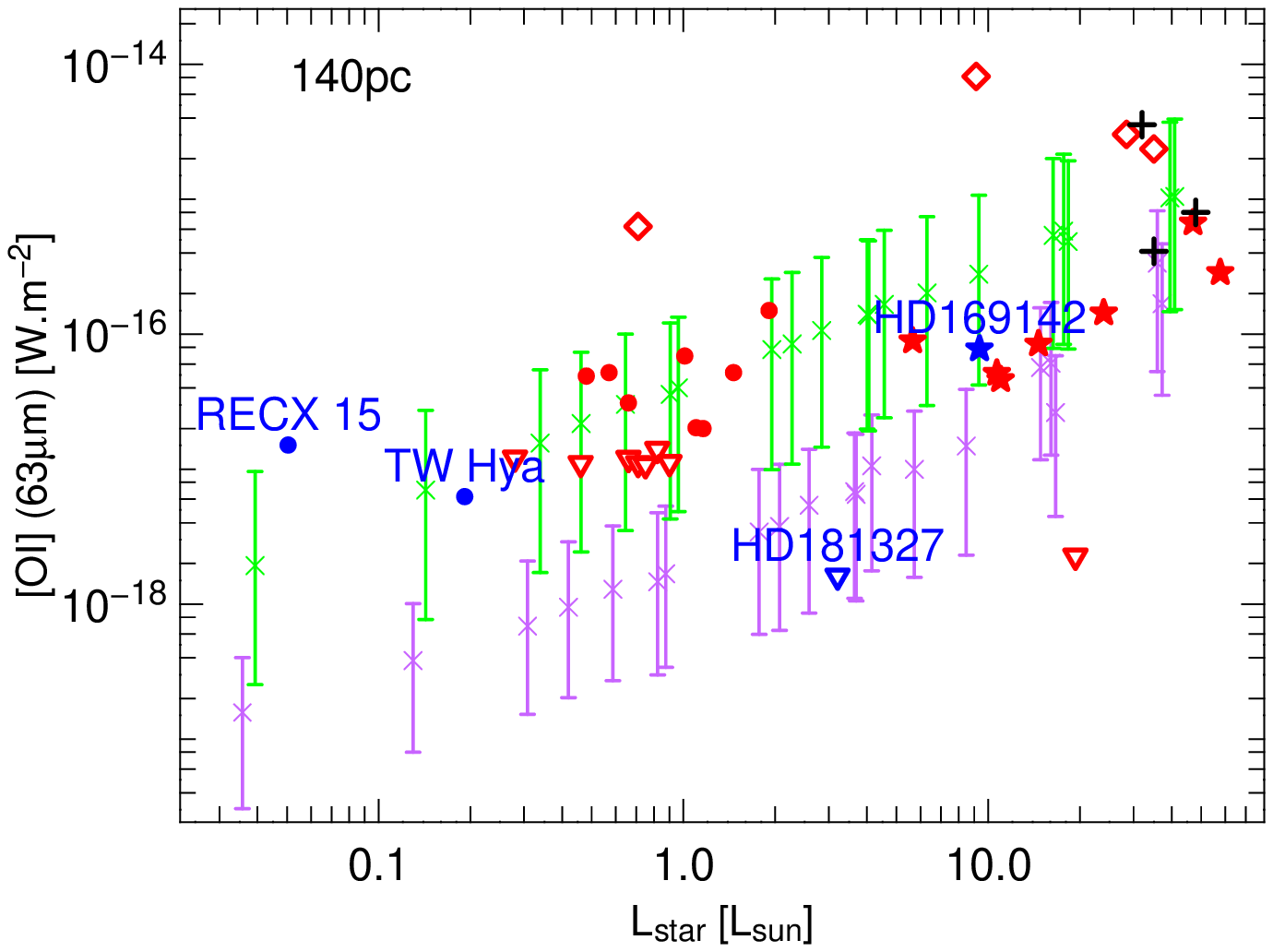}
  \hfill
  \includegraphics[width=0.46\hsize]{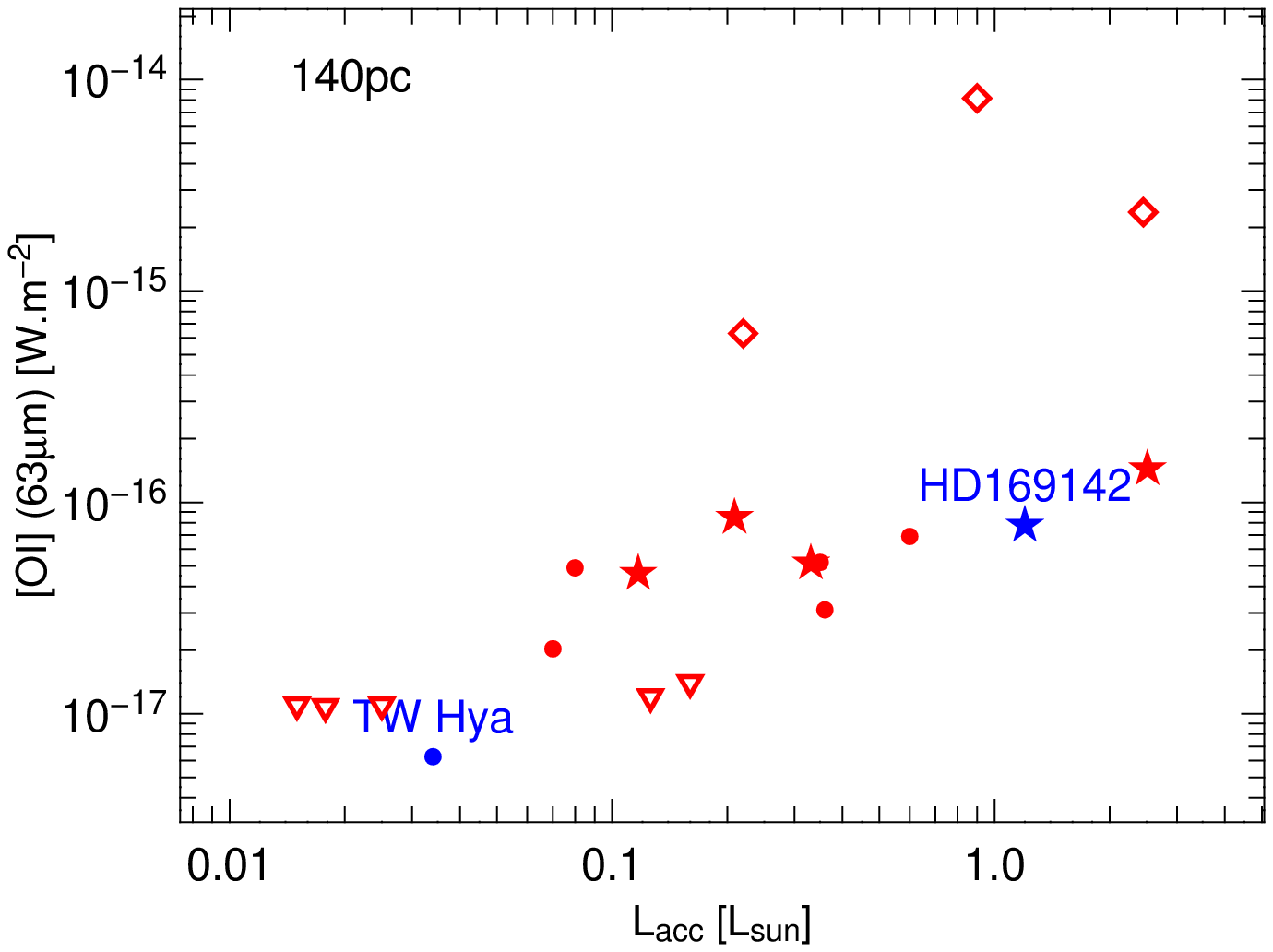}
  \caption{[OI] 63\,$\mu$m line flux as a function of the stellar
    luminosity (left panel) and accretion luminosity (right
    panel). Red and blue
    points, stars, triangles, and diamonds represents GASPS
    observations. Points are T~Tauri stars while stars are Herbig
    stars. Diamonds
    are sources with extended 
    emission, which could originate from an outflow or envelope.
    Triangles are GASPS upper
    limits. Names are only
    indicated for SDP sources (blue).
    Black pluses are ISO observations. 
    Observations were
    scaled to a distance of 140\,pc. The crosses show
    the median values of the DENT models. The full vertical lines
    represent the 1-sigma range of models. 
    Only models with a gas
    mass larger than $10^{-4}$\,M$_\odot$ are plotted. Green and
    purple points 
    are high- and low-UV models respectively (f$_\mathrm{UV} = 0.1$
    and 0.001). Accretion luminosities
    from \cite{Hartmann98} and \cite{Garcia-Lopez06}.
   \label{fig:GASPS1}}
\end{figure*}

\section{The DENT grid and initial GASPS data}

The DENT grid is intended as a statistical tool to investigate the influence
of stellar, disc, and dust properties on the various continuum and line
observables, and to study to what extent these  
dependencies can be inverted to retrieve disc properties. The grid
relies on the combined 
capabilities of the 3D radiative transfer code \mcfost\
\citep{Pinte06,Pinte09} and the gas thermal balance and chemistry code
\ProDiMo\ \citep{Woitke09,Kamp10}. Spectral energy distributions and line fluxes of [OI], [CII],
$^{12}$CO, ortho-H$_2$O and para-H$_2$O are predicted for more than 300\,000 discs
models.
The DENT grid was built by systematically exploring an 11-dimension
parameter space (see \citealp{Woitke10}, table~1). In particular the DENT grid
explores the effect of varying the central star (age, mass, UV excess),
disc dust mass and gas-to-dust mass ratio, inner and outer radii,
flaring and surface density exponents, grain sizes, and presence of
dust settling. 
It is important to keep in
mind that even with the large number of calculated models, the
sampling of each parameter remains coarse and 
that the DENT grid
does not reflect the statistics of objects in GASPS (each
parameter has been sampled uniformly and not following the
distributions of the GASPS target list). 
We refer the reader to
\cite{Woitke10} for details about the grid properties and
computational implementation.

We include here data obtained during the science demonstration
 phase \citep[SDP,][]{Mathews10},
as well as GASPS data reduced prior to 2010
April 23, which will be presented in detail in following papers.
Due to the limited number of sources, statistical analyses remain
premature, but initial comparisons with the 
model predictions are necessary to ensure that the range of models cover the
GASPS observations.

\section{Results and discussion}

One of the main reasons to compute the DENT grid was to estimate the
degeneracies between parameters, i.e. how they
influence the 
various lines and how far \emph{Herschel} line
observations can be inverted to assess the physical and chemical
conditions of the disc.
Not surprisingly, the DENT grid revealed that many parameters affect
the predicted line fluxes and SEDs, and  degeneracies between
parameters are common and complex, 
which makes the interpretation of lines fluxes difficult.

\subsection{Gas heating processes}

\begin{figure*}
  \includegraphics[width=0.46\hsize]{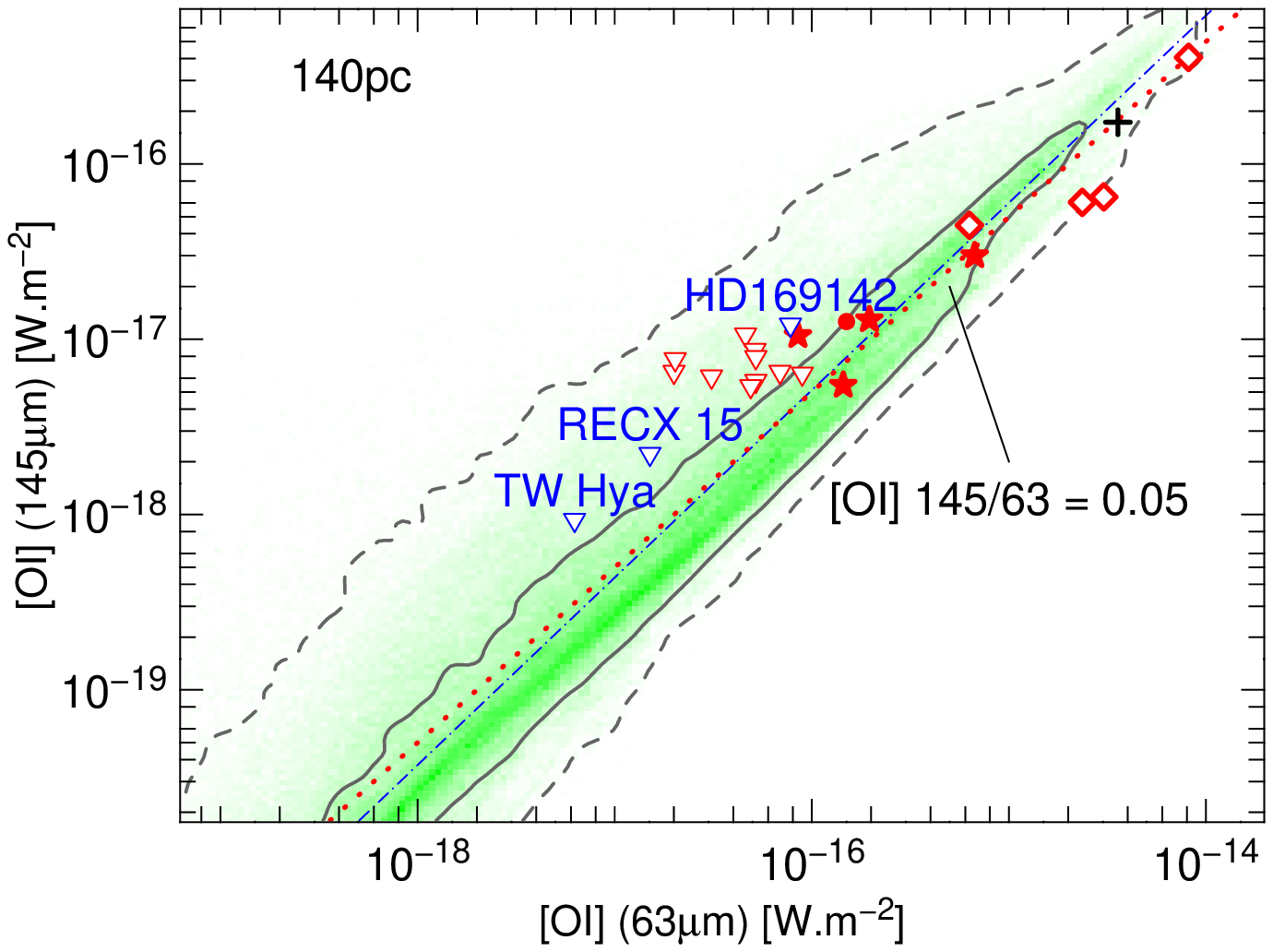}
  \hfill
  \includegraphics[width=0.46\hsize]{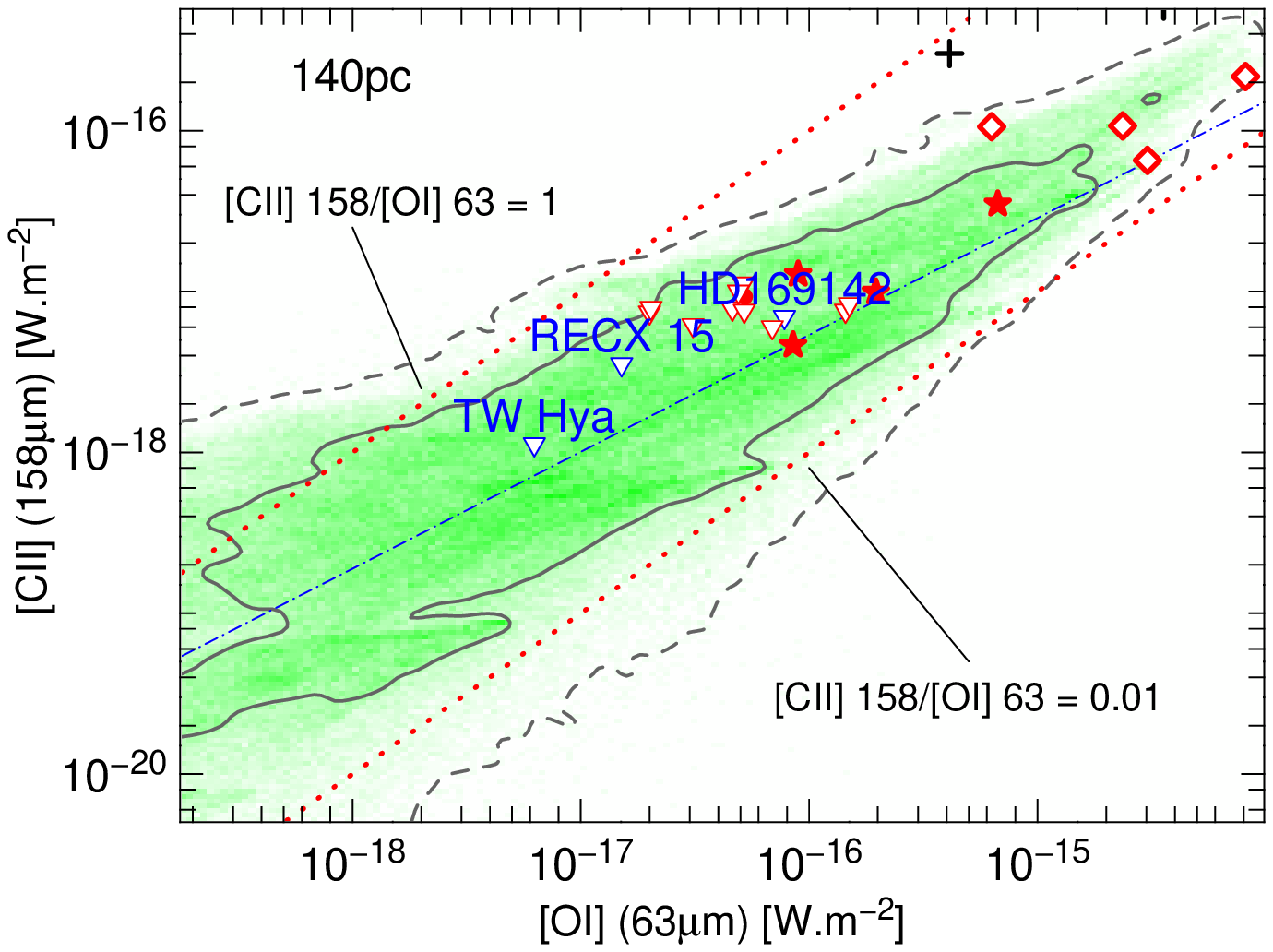}
  \caption{Correlations between line fluxes. Symbols as in
    Fig.\,\ref{fig:GASPS1}. \emph{Left panel:} [OI]
    145\,$\mu$m as a function of [OI] 63\,$\mu$m. 
    \emph{Right panel:}  [CII] 158\,$\mu$m as a function of [OI]
    63\,$\mu$m. Small green points: DENT models. 
    The full and dashed grey contours represent the regions that
    contain 68\,\% and 99.7\,\% of
    the models respectively.
    The blue dot-dashed lines represent power law regression fits
    to the entire DENT grid. The red dotted lines correspond to constant line
    ratios.  In the left panel the line
    corresponds to the average ratio in the DENT grid: 0.05.
    This ratio corresponds to the limit case for
    optically thick lines at temperatures higher than a few hundred
    Kelvin \citep{Tielens85}.\label{fig:line_ratio}}
\end{figure*}

Figure\,\ref{fig:GASPS1} plots the [OI] 63\,$\mu$m line flux as a
function of the stellar luminosity and accretion luminosity.
The DENT grid predicts a correlation  between the line
flux and the stellar luminosity.  All observational points (except
sources with a large outflow or envelope)
 lie within the 1-sigma envelope of the models.

For T~Tauri stars, the detected line fluxes are well reproduced by
models with a high UV excess (f$_\mathrm{UV}$ = 0.1, see \citealp{Woitke10}),
suggesting that UV emission produced by accretion onto the star is one
of the main gas-heating processes. 
 The right panel of
Fig.\,\ref{fig:GASPS1} indeed suggests a trend between [OI] line flux
and accretion luminosity.
On the other hand, Herbig~Ae/Be stars show a strong correlation between the line
flux and stellar luminosities, with a much smaller scatter than for
T~Tauri stars. Large UV excesses do not seem necessary to reproduce
the Herbig observations (data points lie between models with high- and
low-UV excess).
This suggests that stellar radiation is the dominant gas-heating source
for Herbig stars. Because these sources radiate large phostospheric UV
emission, the accretion luminosity represents a smaller fraction of the UV
luminosity and is not as critical  a gas heating mechanism as for
 T~Tauri stars. This is also consistent with the small fraction of
 large accretors (L$_\mathrm{acc} > 0.1\,$L$_*$) among Herbig stars
 \citep{Garcia-Lopez06}.

X-ray irradiation, which
  is not yet included in the DENT grid, can also contribute
significantly to the gas 
heating and chemistry for low-mass objects
\citep[e.g.][]{Glassgold04,Semenov04,Meijerink08,Hollenbach09,Ercolano10} 
and higher fluxes can be expected for
sources with typical T~Tauri X-ray emission. 
The small number of sources observed by GASPS so far prevents us from
distinguishing between UV and 
X-rays for the main heating
process for low mass objects. The full GASPS survey should provide
detailed answers on these aspects.

Figure\,\ref{fig:line_ratio} presents the [OI] 145\,$\mu$m and [CII]
158\,$\mu$m line fluxes as a function of the [OI] 63$\mu$m line. The models
are in excellent agreement with the GASPS observations.
The DENT grid predicts a correlation between the [OI] 63 and
145\,$\mu$m line fluxes. 
A regression fit of all the DENT points indicates that both line fluxes
are almost proportional ($f(\textrm{[OI]}~145\,\mu\textrm{m}) \propto
f(\textrm{[OI]~}63\,\mu\textrm{m})^{0.98}$), with a [OI] 145/63 line
ratio around 0.05 on average. 
The presence of scatter in the plot illustrates the wide range of
physical conditions encountered in the DENT grid. The correlation, however,
is very strong (Pearson correlation coefficient of 0.97) and holds for several
orders of magnitude in line flux. 
These results agree with the prediction of \cite{Tielens85}
for a 1D photodissociation region (see their Fig.\,2).
Our detailed line modelling of the 2D PDR disc surface with varying
density and irradiation confirms the picture drawn from these 1D
models: the oxygen lines are optically
thick and originate in
a relatively high-temperature gas ($\geq 100\,$K).
 As a
consequence, this line ratio does not provide
constraints on the local gas density and temperature in most cases.

Deviations from this average ratio of 0.05 are interesting though.
In particular, two Herbig~Ae with known outflows observed by GASPS present a small line
ratio around 0.025. According to \cite{Tielens85}, this value cannot
be obtained for optically thick lines. This suggests that a
significant fraction of the line fluxes originates
from an optically thin region  above the bulk of the disc (potentially
the outflow) with a
temperature between 40 and 200\,K. These sources will be studied in
detail in following GASPS papers.

The [CII] 158\,$\mu$m line also presents a correlation with the [OI]
63\,$\mu$m, but with a much larger scatter. A regression fit of the
DENT models indicates
that on average $f(\textrm{[CII]~}158\,\mu\textrm{m}) \propto
f(\textrm{[OI]~}63\,\mu\textrm{m})^{0.7}$, with a decreasing line ratio as
the line flux increases. This suggests an increasing gas temperature
(Fig.\,3 in \citealp{Tielens85}) in the disc towards
more luminous objects. Most of the DENT models lie in a region where
the line ratio is between 0.01 and 1, suggesting gas temperatures higher
than 100\,K. More detailed analyses of the [CII] line
 are complicated by several
factors: our disc models show that the line originates from larger radii
and lower density regions than the [OI] lines, and it is very sensitive to the
amount of UV radiation. 

\subsection{Gas mass and gas-to-dust mass ratio}

\begin{figure*}
  \centering
  \includegraphics[width=0.46\hsize]{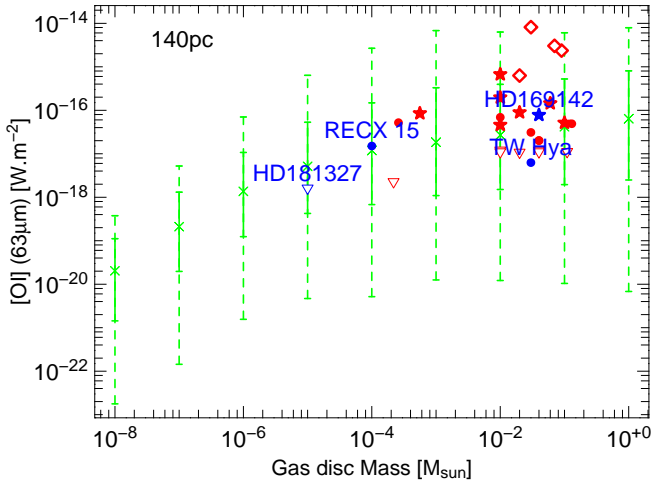}
  \hfill
  \includegraphics[width=0.46\hsize]{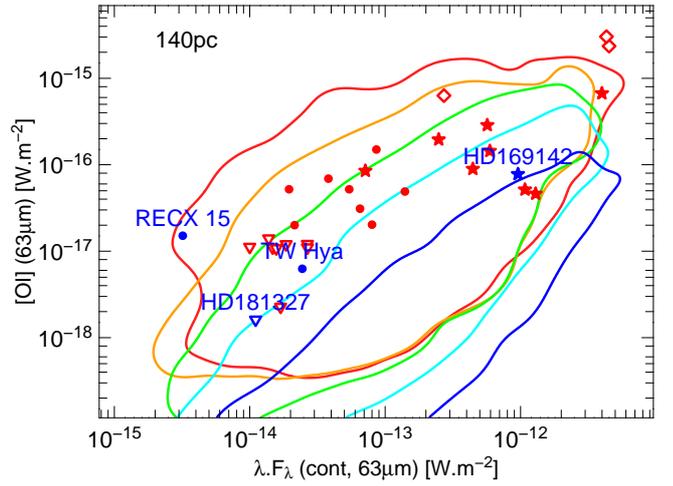}
  \caption{[OI] 63\,$\mu$m line flux as a function of the gas mass
    (left) and adjacent
    continuum (right). Symbols as in Fig.\,\ref{fig:GASPS1}.  The full and dashed vertical lines
    represent the 1- and 3-sigma range of models. 
    The values for the gas mass of the observed sources
    are \emph{indicative} only, they were estimated from millimetre emission
    and assuming a gas-to-dust ratio of 100. On the right panel,
    contours represent the regions that contain 68\,\% of
    the models as a function of the gas-to-dust mass ratio: red = 1000,
    orange = 100, green = 10, light blue = 1, dark blue = 0.1.  \label{fig:line_cont}}
\end{figure*}

In addition to the main trend of an increasing line flux with
  (UV) luminosity,  
there is also a trend with gas mass,
where the  synthetic line flux increases with mass,  
but the correlation seems to saturate above $10^{-4}$\,M$_\odot$
\citep[see also][]{Woitke10}, preventing direct inversion of the line
flux into a gas mass (Fig.\,\ref{fig:line_cont}).

The right panel of Fig.\,\ref{fig:line_cont} shows the correlation between the [OI]
63\,$\mu$m line flux and the adjacent continuum.
Because the line and continuum emissions are optically thick in
most cases, these fluxes give an indication of the relative
temperatures and projected surface area of the emitting regions (gas and dust). As the
stellar luminosity increases, the region of the disc which is warm enough to
contribute significantly to the emission also increases, resulting in
larger fluxes.
This behaviour is observed for most GASPS sources where the line flux
roughly increases with the continuum level, but with a significant
  scatter. As a consequence, this
indicates that a large fraction of discs with a significant far-IR
excess will be detected in [OI] by \emph{Herschel}.
The separation of the DENT models according to their
gas-to-dust mass ratio 
suggests that most objects are gas-rich (gas/dust mass ratio~$>10$). 
The large scatter in
the models, and the optical depth in the continuum and the line,
precludes however,
 in most cases, a precise estimate of the gas-to-dust
ratio for individual sources. For
instance, no direct ratio (or upper limit for HD~181327) can be estimated from this diagnostic alone for
 HD~169142 and  TW~Hydra, 
for which the  
line fluxes can be reproduced by any ratio between 1 and 1\,000.
In addition, the contribution of an
  outflow to the line flux  may affect the estimation of the disc gas-to-dust 
ratio and needs to be accounted for.

\begin{figure}
  \centering
  \includegraphics[width=0.92\hsize]{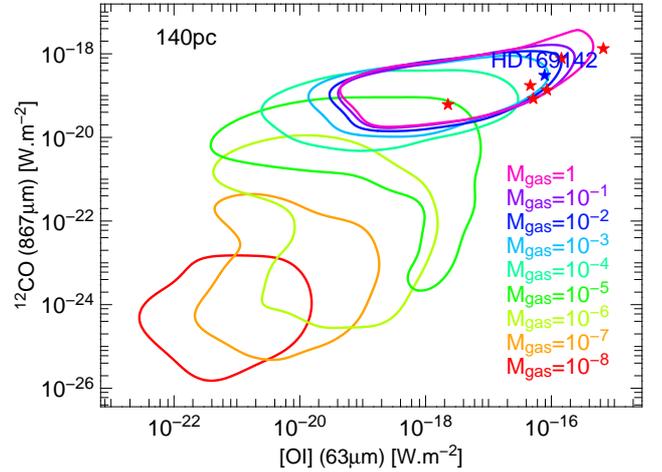}
  \caption{$^{12}$CO J=3$\rightarrow$2 (867\,$\mu$m) as a function of [OI]~63\,$\mu$m.
     Contours represent the
    regions that contain 68\,\% of 
    the models as a function of the gas mass (in solar masses). Data
    points from \cite{Dent05}.
     \label{fig:CO_OI}}
\end{figure}

Greater observational constraints 
and more detailed modelling is
required to estimate the gas mass and gas-to-dust 
ratio.  In particular,
the combination of low rotational level transitions of CO with oxygen
lines offers a valuable proxy to estimate the amount of gas in
discs. 
Figure\,\ref{fig:CO_OI} plots the $^{12}$CO~J=3$\rightarrow$2
line flux
as a function of the [OI] 63\,$\mu$m line flux. We stress that the accuracy in the calculated CO abundances
is limited by our approximate treatment of self-shielding (see
\citealp{Woitke09}), but this does not affect our conclusions. For low-mass
discs, this diagram allows a clear distinction of the gas disc
mass.
As the mass increases, lines become optically thick and the
corresponding fluxes saturate, preventing determination of the gas
mass. Current CO surveys  can only reach sources in this
  saturation regime (see for instance data from
  \citealp{Dent05} in Fig.\,\ref{fig:CO_OI}), but
this perspective is particularly interesting in the context of
\emph{ALMA}, which will offer high sensitivity for CO lines ($\approx 10^{-23}$\,W.m$^{-2}$).
Similar diagrams combining $^{13}$CO (not included in DENT, but see
  \citealp{Meeus10} and \citealp{Thi10}),
C$^{18}$O and [OI] 145\,$\mu$m, which saturate at higher masses
due to lower optical depths, will further help to overcome this
degeneracy. As oxygen lines are sensitive to warm
gas in the inner 10-30\,AU (for T~Tauri stars), they offer
complementary views to the low-J CO lines which probe regions outside
of 20-40\, AU, especially
 when resolved maps
of the CO emission are available.

\section{Summary and conclusions}

The GASPS survey will offer unique views of the gas and dust phases of
protoplanetary discs. In order to provide statistical tools to help the
interpretation of the survey results, we interfaced the \mcfost\
and \ProDiMo\ codes and calculated a large grid
of models sampling the range of discs observed by GASPS. This 
will allow us to determine some of the physical conditions within
discs. 
The initial results from the GASPS survey tend to confirm
the predictions of the DENT grid,
 illustrating
the main parameters affecting the line fluxes, namely the UV excess
and/or X-ray emission for T~Tauri stars and the UV stellar
irradiation for Herbig stars. This is a highly relevant point to be
considered in subsequent open time programs on discs.

The interpretation of line results remains difficult and their inversion
into physical parameters must be performed
with caution, because the DENT grid highlights considerable degeneracies between
parameters and the complex interplay between various physical processes.
The [OI] 63$\,\mu$m is crucial for breaking some of the degeneracies. By
combining this line with continuum and/or (sub)mm rotational lines, we can
possibly distinguish various parameters.
\cite{Meeus10} and \cite{Thi10} illustrate how far this inversion
can be performed when high quality data sets with a wide range of
observational techniques are available.

\begin{acknowledgements}
C. Pinte acknowledges funding from the 
European Commission's 7$^\mathrm{th}$ Framework Program 
as a Marie Curie Intra-European Fellow
(PIEF-GA-2008-220891). 
The members of LAOG, Grenoble acknowledge
 PNPS, CNES and ANR (contract ANR-07-BLAN-0221) for financial
 support. W.F.~Thi acknowledges a SUPA astrobiology fellowship.
G.~Meeus, C.~Eiroa, I.~Mendigut\'ia and B.~Montesinos are partly
supported by Spanish grant AYA 2008-01727. D.R.~Ardila, S.D.~Brittain,
W.~Danchi, 
C.A.~Grady, C.D.~Howard, G.S.~Mathews, I.~Pascucci, A.~Roberge,
B.~Riaz, G.~Sandell and J.P.~Williams acknowledge NASA/JPL for
funding support. J.M.~Alcid and E.~Solano acknowledges funding from the Spanish
MICINN (grant AYA2008-02156). 
\end{acknowledgements}

\bibliography{biblio}

\end{document}